DRAFT

# Pump-probe polarized transient hole burning (PTHB) dynamics of hydrated electron revisited. [1]

Ilya A. Shkrob [(1)]

*Chemistry Division , Argonne National Laboratory, Argonne, IL 60439*

Ross E. Larsen, William J. Glover, Benjamin J. Schwartz [(2)]

*Department of Chemistry and Biochemistry, University of California,*

*Los Angeles, CA 90095-1569*

**Abstract**

Femtosecond PTHB spectroscopy was expected to demonstrate the existence of distinct *s*-*p* absorption subbands originating from the three nondegenerate *p*-like excited states of hydrated electron in anisotropic solvation cavity. Yet no conclusive experimental evidence either for this subband structure or the reorientation of the cavity on the picosecond time scale has been obtained. We demonstrate that rapid reorientation of *s*-*p* transition dipole moments in response to small scale motion of water molecules is the likely culprit. The polarized bleach is shown to be too small and too short lived to be observed reliably on the sub-picosecond time scale.

---

[*] Authors to whom correspondence should be addressed; electronic mail: (1) shkrob@anl.gov, (2) schwartz@chem.ucla.edu,

## 1. Introduction

The current dynamic models for hydrated electron, $e^-_{hyd}$, [1-11] suggest that anisotropy of the solvation cavity should result in significant energy separation between the three *p*-like excited states of the $e^-_{hyd}$ and the presence of three overlapping *s-p* subbands in the absorption spectrum. Yet the only incontrovertible evidence indicating this structure is from the excitation energy dependence for depolarization ratio of the O-H stretch in stimulated Raman spectra of the $e^-_{hyd}$. [12] Laser experiments [13-17] that were specifically designed to demonstrate this subband structure using polarized transient hole burning (PTHB) spectroscopy yielded no conclusive evidence for this predicted structure. In this Letter, we examine the probable cause for this negative result.

The PTHB spectroscopy is a form of ultrafast pump-probe spectroscopy [18] that examines the *ground-state* dynamics of a system by first exciting a subset of members of an ensemble with polarized light and then probing at a later time the transient absorption dynamics of the remaining, unexcited members with probe light polarized parallel or perpendicular to the polarization of the pump. [13-19] Cavanagh *et al*. [17] have recently discussed the application of PTHB to the $e^-_{hyd}$ in great detail, so here we give only a brief outline. If the three *p*-like states $p_x$, $p_y$, and $p_z$ interchange roles slowly (as the cavity deforms in response to solvent motion), then the PTHB should show different dynamics for parallel and perpendicular probe polarizations. When the lowest-energy *s-p* transition (along the long axis *z* of the solvation cavity) is excited with polarized light, until the cavity axes reorient, there would be *less* $s$-$p_z$ absorption by the remaining electrons when probing with light of the same polarization at the excitation energy, but the remaining electrons would continue to absorb at higher energies, in their $s$-$p_x$ and $s$-$p_y$ subbands. If the probe light were polarized perpendicular to the exciting light, the opposite trend would be observed. Mixed quantum-classical molecular dynamics (MQC MD) simulations of Schwartz and Rossky [19] predicted that pumping the lowest-lying $s$-$p_z$ transition should yield PTHB anisotropy that persists for ≈1 ps, as it takes this long for the cavity to change shape and/or orientation. [20] While this anisotropy was initially observed by Reid *et al.*, [13] the subsequent studies did not confirm their result, [14-17] suggesting the rapid loss of orientational memory.

In this Letter, we revisit the PTHB dynamics theoretically, by using two complementary approaches. The first approach is based on the use of mobile (floating) sets of Gaussian orbitals [7-9,22] (FSGO). The second approach is based on configuration interaction with single excitations (CIS) for dynamically embedded water anion clusters. [11] Both of these models used the same MQC MD trajectory as their input and expanded the electron wavefunctions locally. The FSGO model involves the use of *ad hoc* pseudopotential for interaction of the excess electron with water molecules (which allowed us to include several solvation shells) whereas CIS model treats explicitly only 15-25 water molecules forming the solvation cavity (but at *ab initio* level). [11] As shown below, both of these models make predictions of rapid (< 100 fs) depolarization.

## 2. Computational Details

The input for these calculations was provided from a 100-ps adiabatic MQC MD trajectory with a time step of 1 fs. See ref. [11] for more detail of this calculation. In the FSGO model, the electron wavefunction $\phi$ is expanded on a set of primitive Gaussian orbitals (GO's) of the type $\chi(r) \propto \exp\left[-f\alpha r^2\right]$ arranged in *s*-, *p*-, and *d*-shells that are centered on the center of mass (*X*), which was adjusted iteratively. The cavity-centered basis set (V4) consisted of four *s/p, s/p/d, s/p/d,* and *s*- shells with $\alpha$=0.018 Å$^{-2}$ and $f$=20, 10, 2, and 1, respectively. To this set we added (water-centered) sets (W2) that included O 2*s/p* and H 2*s* orbitals (with $f$=1 and $\alpha$=1.762 and 0.833 Å$^{-2}$, respectively, which correspond to the outer orbitals of the standard STO-2G set [21]). The main effect of this extension was in the compression of the electron wavefunction inside the cavity due to the exclusion by water molecules. The interaction Hamiltonian is given by pairwise additive screened Coulomb potential. [8] The ground state wavefunction $\phi$ was used to calculate the gyration tensor $\mathbf{G}_{ik} = \langle \phi | \mathbf{x}_i \mathbf{x}_k | \phi \rangle$ with principal semiaxes $r_a < r_b < r_c$; the difference between these semiaxes is a measure of anisotropy of the solvation cavity. The gyration radius $r_g$ for the ground state electron is defined as $r_g^2 = tr(\mathbf{G})$. In our CIS calculations, a 6-31G split-valence double-ζ Gaussian basis set augmented with diffuse and polarized (d,p) functions was used. [11,21] Water molecules of one or two complete solvation shells were treated explicitly, and the remaining

“matrix” molecules were replaced by point fractional charges. [11] The calculations included the first 10 excited states, which were used to obtain transition dipole moments $\boldsymbol{\mu}_{0i}$ and transition energies $E_{0i} = \hbar\omega_{0i}$. While the absorption spectra shift to the blue as more water molecules are included, the calculated autocorrelation functions and PTHB dynamics do not change significantly as more solvent shells are added.

For a pump beam of frequency $\omega_{ex}$, the changes $\Delta OD_{\parallel,\perp}(\omega,t)$ in the optical density at the delay time *t* after photoexcitation, for probe light with frequency $\omega$ that is polarized parallel (||) and perpendicular ($\perp$) with respect to the excitation pulse, is obtained from

$$\Delta OD(\omega,t) \propto \left\langle \sum_{i,j=1}^{3} \left|\boldsymbol{\mu}_{0i}(t)\right|^2 \left|\boldsymbol{\mu}_{0j}(0)\right|^2 \ \Gamma_{ij}(\omega,\omega_{ex};t) \right\rangle, \quad (1)$$

$$\Delta A(\omega,t) \propto \frac{2}{5} \left\langle \sum_{i,j=1}^{3} \Xi_{ij}(t) \left|\boldsymbol{\mu}_{0i}(t)\right|^2 \left|\boldsymbol{\mu}_{0j}(0)\right|^2 \Gamma_{ij}(\omega,\omega_{ex};t) \right\rangle . \quad (2)$$

The latter expressions are given for the isotropic ($\Delta OD = \Delta OD_{\parallel} + 2\Delta OD_{\perp}$) and anisotropic ($\Delta A = \Delta OD_{\parallel} - \Delta OD_{\perp}$) contributions averaged over all possible orientations with respect to the laboratory frame, where

$$\Gamma_{ij}(\omega,\omega_{pump};t) = \omega\delta(\omega - \omega_{0i}(t)) \times \omega_{ex}\delta(\omega_{ex} - \omega_{0j}(0)) \quad (3)$$

and

$$\Xi_{ij}(t) = \left(3\left|\hat{\boldsymbol{\mu}}_{0i}(t) \bullet \hat{\boldsymbol{\mu}}_{0j}(0)\right|^2 - 1\right)\Big/2$$

(the roof indicates a unit vector in the direction of $\boldsymbol{\mu}_{0i}$). In eqs. 1 and 2, $\boldsymbol{\mu}_{oi}(t)$ is the electronic transition dipole moment between the ground state and the excited states (in the order of increasing energy) and the angled brackets denote an ensemble average. No vibronic transitions are included and the absorption of electronic excited states in the spectral region of interest is neglected. The PTHB dynamics may be characterized by autocorrelation functions $C_i(\tau) = \langle \Xi_{ii}(\tau) \rangle$ that provide a measure of the memory loss for the direction of the transition dipole. The spectral diffusion is characterized by the autocorrelation function $E_i(\tau) = \langle \delta\omega_{0i}(\tau)\delta\omega_{0i}(0) \rangle / \langle \omega_{0i} \rangle^2$, where $\delta\omega_{0i}(\tau) = \omega_{0i}(\tau) - \langle \omega_{0i} \rangle$. In our numerical calculations, the excitation and probe bandwidths were set to 0.1 eV.

## 3. Results and Discussion

The absorption spectrum calculated for the $e^-_{hyd}$ using the FSGO model (Figure 1a) exhibits the maximum at 1.8-2 eV (vs. the experimental 1.72 eV at 300 K). [22] The estimate $r_g^2 \approx 3\hbar^2 / 2m_e \langle E^{-1} \rangle$ for the gyration radius $r_g$ obtained from this spectrum is 2.41 Å (vs. experimental 2.5-2.6 Å ); [22] direct calculation gives $r_g$ =2.64 Å. The solvation cavity is highly anisotropic: the mean eccentricities [11] of the gyration ellipsoid are ≈0.5. This anisotropy splits the levels of the *p*-like states; the principal axes of the gyration tensor roughly correspond to the directions of transition dipole moments for the *p*-like states. It is seen from Figure 1b that there is significant overlap of the *s*-*p* subbands that facilitates spectral diffusion. Figure 2 demonstrates calculated autocorrelation functions. While there is a slow component in these autocorrelation functions (with a life time of 1.5-2 ps), there is also a fast component that accounts for most of the decrease. For transitions involving the highest and the lowest *p*-like states, ca. 75% of the transition moments decorrelate in 100 fs; for the middle *p*-like state, this loss amounts to 95% (Figure 2a).

In Figure 3a, we calculate isotropic and anisotropic components of PTHB spectra for $\hbar\omega_{ex}$ =1.8 eV. As the number of states used to calculate PTHB spectra increases in proportion to the overall absorption, pump excitation near the center of the absorption band offers better statistics (although it reduces the polarized THB signal). It is clear from Figure 3a that even at the delay time of 100 fs, the anisotropy ($\Delta A$) is only 5% of the isotropic one ($\Delta OD$). (Note the scaling factor of 10 for $\Delta A(\omega, t)$). By *t*=300 fs, the anisotropy is further reduced by a factor of two. As the anisotropy is maximum when a single *s*-*p* transition is photoexcited, in Figure 3b we calculated the PTHB spectra for excitation at the red ($\hbar\omega_{ex}$=1.5 eV) and blue ($\hbar\omega_{ex}$=2.2 eV) edges of the absorption spectrum. Even under such favorable excitation conditions, the anisotropic contribution is 2-5% of the isotropic one. The decay kinetics of the $\hbar\omega_{ex}$=1.5 eV photon induced $\Delta A$ signals at $\hbar\omega$=1.5 eV (bleach) and $\hbar\omega$=1.9 eV (absorption) are shown in Figure 3c. Both of these signals decay with a time constant of 100-150 fs. The smallness of the PTHB anisotropy and its rapid decay readily account for the difficulty of experimental detection against the ever present coherent artifact due to induced birefringence from the optical Kerr effect. [17]

The loss of the memory is actually *faster* than 100 fs. Figure 1S in the supplement shows autocorrelation function $C_i(\tau)$ obtained using snapshots taken every 20 fs along the same MQC MD trajectory. It is seen from this plot that most of decorrelation occurs in < 40 fs; the time constants for this decorrelation are 20-50 fs. The decorrelation is so rapid that resolving it in time experimentally would require using very short excitation pulses (5-10 fs). As such pulses do not excite the *s-p* subbands selectively, no clear distinction can be made between the different orientations, so the PTHB experiment with a better time resolution would not provide significant improvement. Figure 2S shows the autocorrelation functions obtained using CIS calculations for embedded water anion clusters. Despite the difference in the approach, these autocorrelation functions are very similar to those shown in Figure 2. The calculated PTHB dynamics shown in Figure 3S are also qualitatively similar to the dynamics shown in Figure 3.

**4. Conclusion.**

The calculated PTHB spectra show rapid (< 100 fs) decorrelation of transition dipole moments for $e^-_{hyd}$. The slow component (which corresponds to reorientation of the cavity as a whole) observed in calculations of Schwartz and Rossky [19] is also observed in these calculations, but this component has low weight, so the calculated anisotropy $\Delta A/\Delta OD$ is at best a few percent, even at the shortest experimentally available delay times (200-300 fs). Such a component would be easily obscured by the coherent artifact from the pump pulse. The root cause for the observed behavior is that the electron wavefunction constantly changes in response to small molecular motions. Following the effect of these rapid changes on the directions of transition dipole moments requires high-quality orbital maps that are provided by local expansion of these wavefunctions. A similar mechanism could explain the lack of persistent anisotropy for solvated electrons in methanol, despite much slower solvent dynamics in this liquid. [17]

## 5. Acknowledgement.

The work at Argonne was supported by the Office of Science, Division of Chemical Sciences, US-DOE under contract No. DE-AC-02-06CH11357. BJS, WJG and REL gratefully acknowledge the support of the National Science Foundation under grant number CHE-0603776.

**References.**

**Figure captions.**

**Figure 1.**

(a) Histogram of the oscillator strengths (the absorption spectrum) of hydrated electron averaged over 1000 snapshots taken from a 100 ps MQC MD trajectory at 296 K. The spectra were calculated using FSGO method with (i) V4 and (ii) V4+W2 basis sets. (b) One-electron DOS function in the FSGO (V4+W2) model (gray dotted line); also shown are histograms of the orbital energies for the ground ("0") and the first five excited states (numbered 1 through 5 in the order of increasing energy). The *s*-, *p*-, and *d*-bands in the DOS function are clearly distinguishable.

**Figure 3.**

Autocorrelation functions (a) $C_k(\tau)$ and (b) $E_k(\tau)$ for the three lowest (p-like) excited states ($k$=1,2,3) of hydrated electron (FSGO model). The solid lines in (a) are biexponential fits.

**Figure 4.**

(a,b) The calculated PTHB spectra for (a) 1.8 eV and (b) 1.5 and 2.2 eV pump excitation of hydrated electron (FSGO model). Note that $\Delta A$ is scaled by a factor of 10 to facilitate the comparison. The dotted line is a scaled absorption spectrum. In panel (a), the anisotropic component for delay times of $t = 100$, 200, and 300 fs are plotted together. In panel (b), only $t$=100 fs spectra are shown. (c) The decay of the anisotropic component following 1.5 eV photoexcitation observed at 1.5 eV (filled circles) and 1.9 eV (empty circles) that correspond to the extrema of the $\Delta A(\omega, t)$ spectrum. The solid lines are exponential fits.

Figure 1

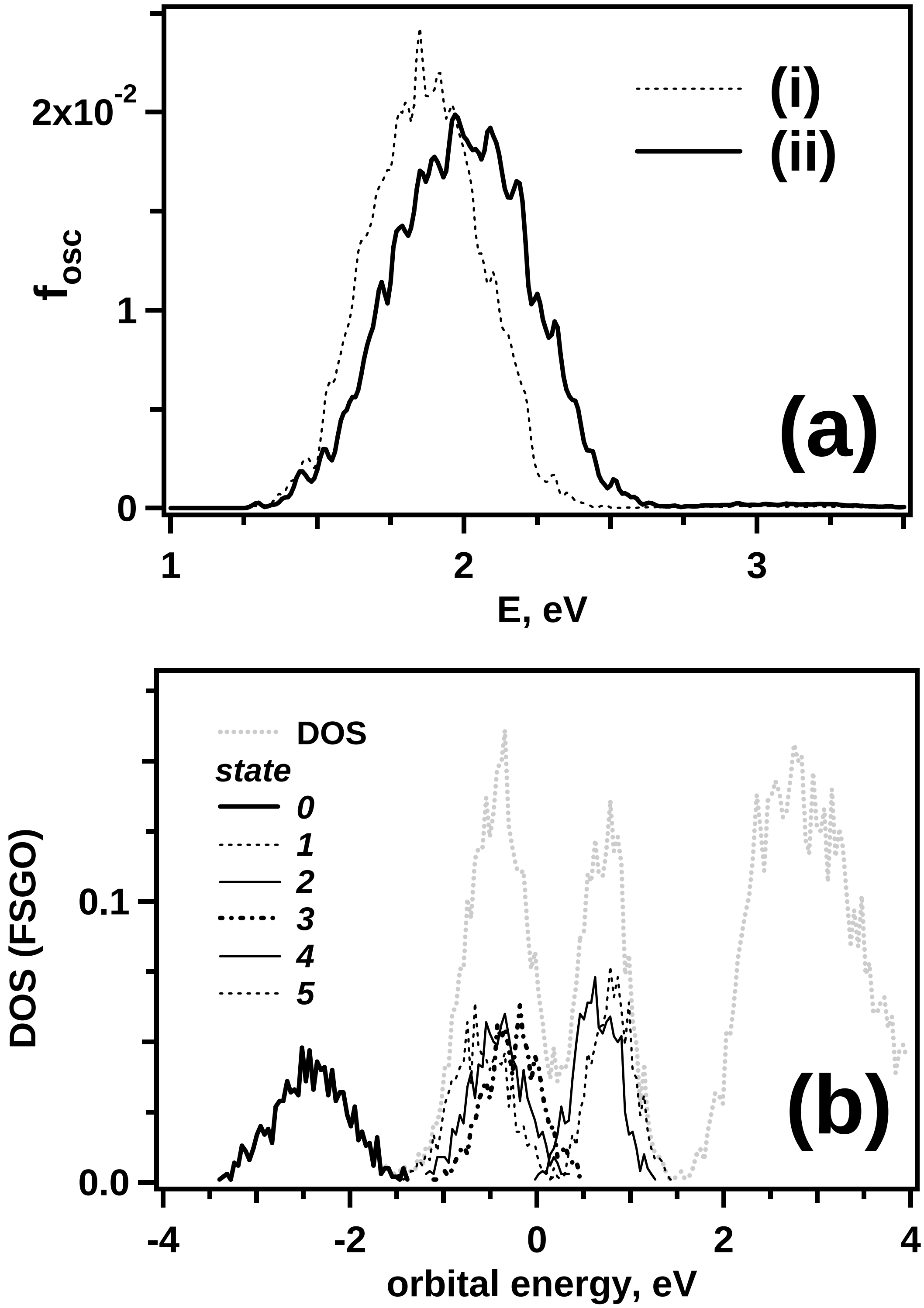

(i)
(ii)
(a)
$f_{osc}$
2x10-2
1
0
1
2
3
E, eV
DOS
state
0
1
2
3
4
5
(b)
DOS (FSGO)
0.1
0.0
-4
-2
0
2
4
orbital energy, eV

Figure 2

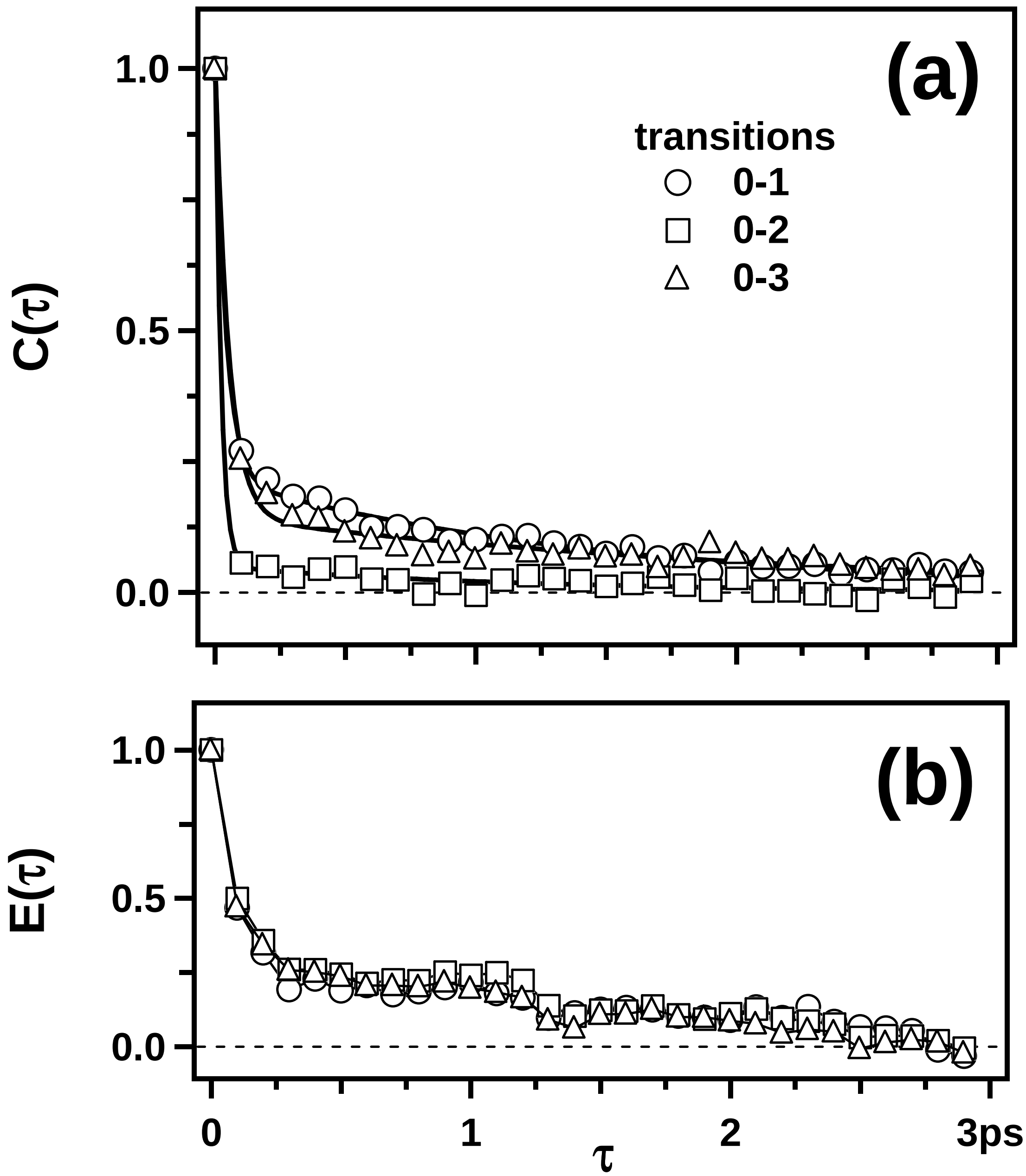

Figure 3

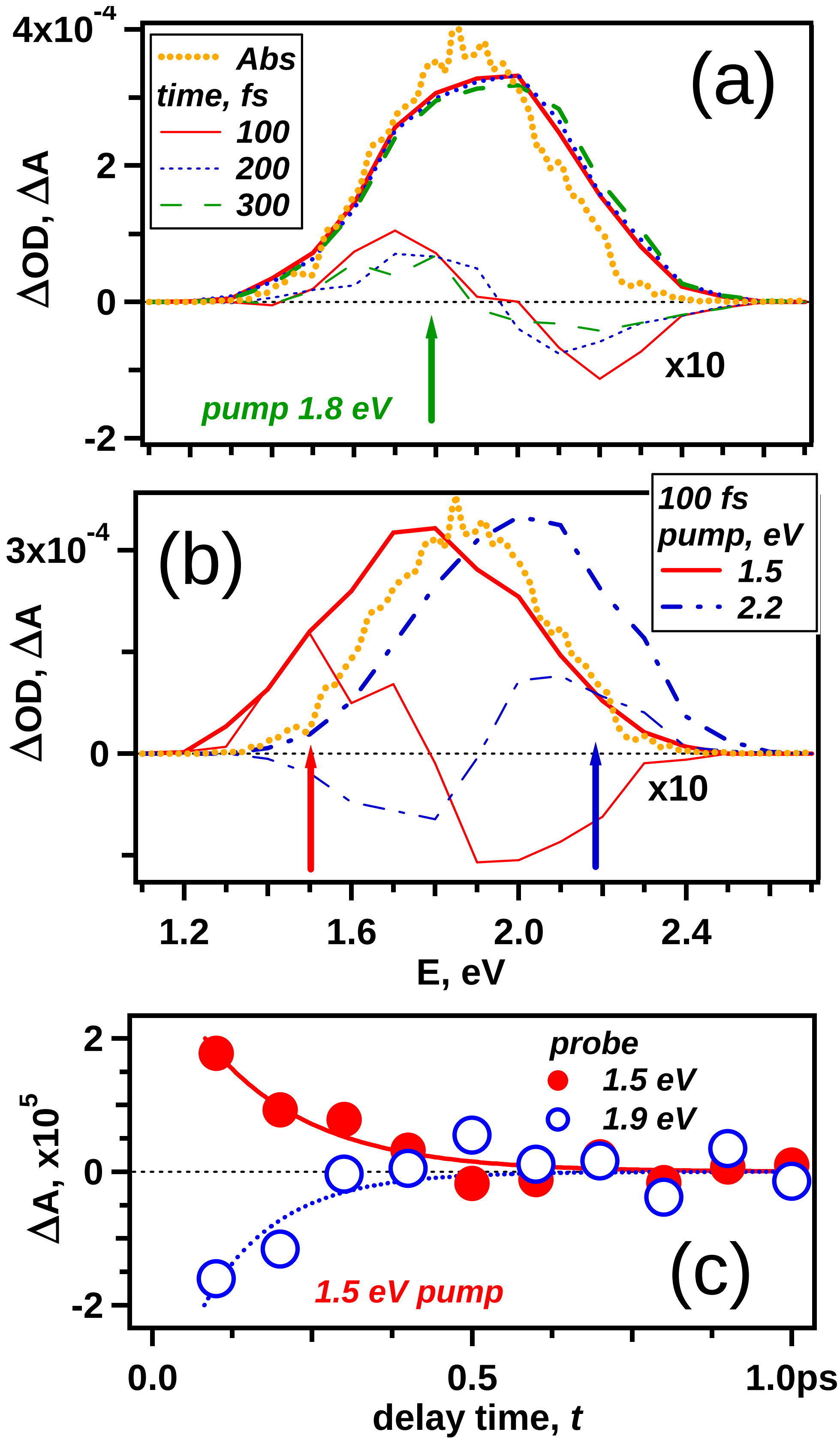

4x10-4
2
0
-2
ΔOD, ΔA
Abs
time, fs
100
200
300
(a)
x10
pump 1.8 eV
3x10-4
0
ΔOD, ΔA
(b)
100 fs
pump, eV
1.5
2.2
x10
1.2
1.6
2.0
2.4
E, eV
2
0
-2
ΔA, x10^5
probe
1.5 eV
1.9 eV
(c)
1.5 eV pump
0.0
0.5
1.0ps
delay time, t

# Supplement

## Pump-probe polarized transient hole burning (PTHB) dynamics of hydrated electron revisited.

by Ilya A. Shkrob (shkrob@anl.gov, Chemistry Division , Argonne National Laboratory, Argonne, IL 6043), Ross E. Larsen, William J. Glover, and Benjamin J. Schwartz (schwartz@chem.ucla.edu, Department of Chemistry and Biochemistry, University of California, Los Angeles, CA 90095-1569)

**Figure 1S.**

As Figure 2a, for (empty symbols) 1000 x 100 fs and (filled symbols) 200 x 20 fs MQC MD trajectory.

**Figure 2S.**

(a,b) As Figure 2, for CIS($N$=10)/6-31+G* calculation involving two complete solvation shells around the electron cavity (embedded in a matrix of fractional charges [11]), for the same MQC MD trajectory. The absorption spectrum is given in the inset, panel (c). Despite the difference in the two methods, the two sets of autocorrelation functions are strikingly similar.

**Figure 3S.**

(a) The same as Figure 3, for polarized THB spectra calculated using the CIS($N$=10)/6-31G* model. For $\hbar\omega_{ex} = 2.0$ eV excitation, $\Delta A$ for $t$=100 and 200 fs are shown. For $\hbar\omega_{ex} = 1.9$ eV excitation, only one delay time, $t$=100 fs is shown. The absorption spectrum is shown by a dotted line. (b) Time evolution of the $\Delta A(\omega,t)$ signal for $\hbar\omega_{ex} = 1.9$ eV photoexcitation at the extrema of the spectrum shown in (a). The filled circles are for $\hbar\omega = 2$ eV probe and the open circles are for 2.7 eV probe. Qualitatively, there are no differences between Figures 3a and 3S, despite the difference in the methods used to obtain the transition dipole moments and energies.

Figure 1S

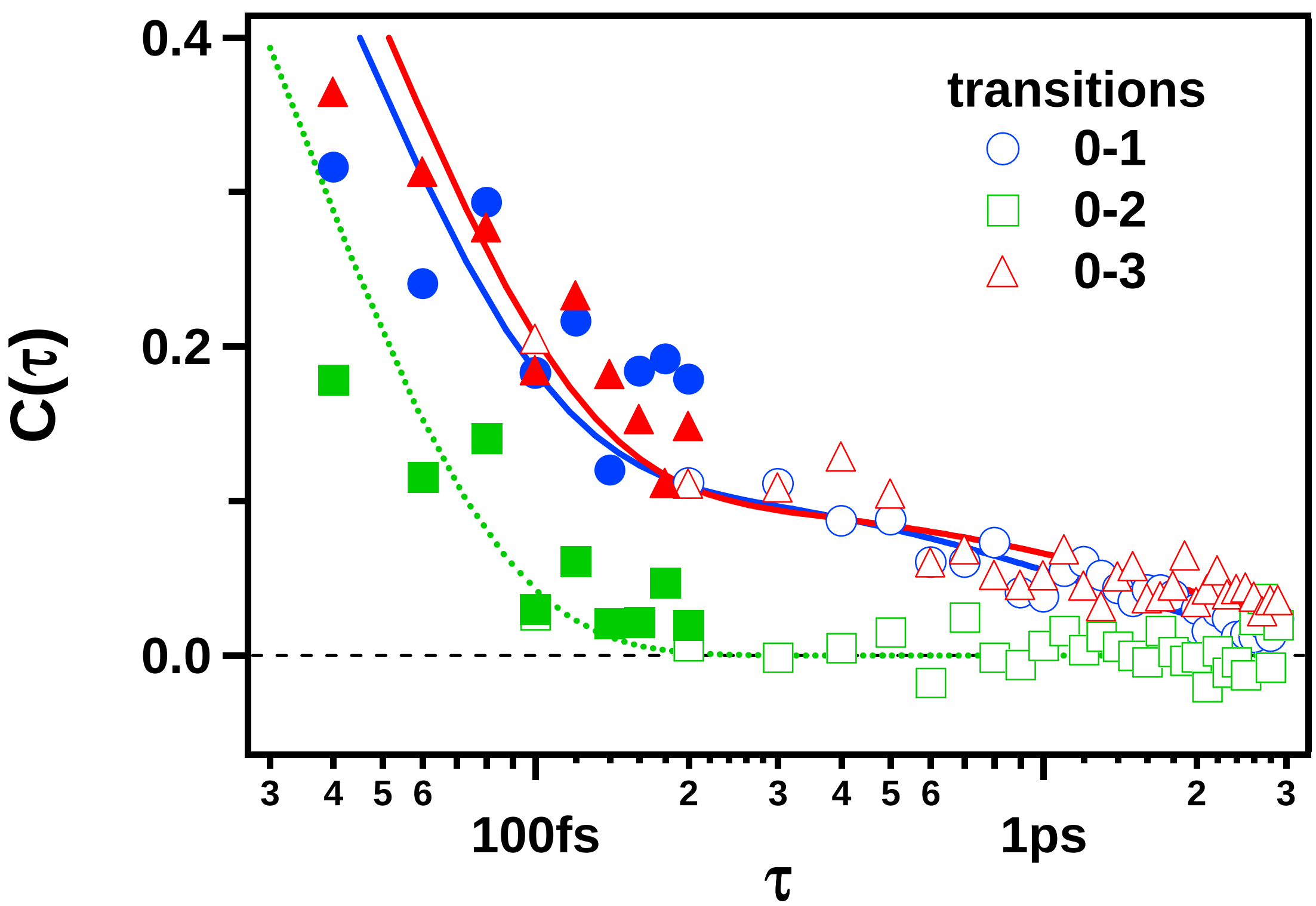

Figure 2S

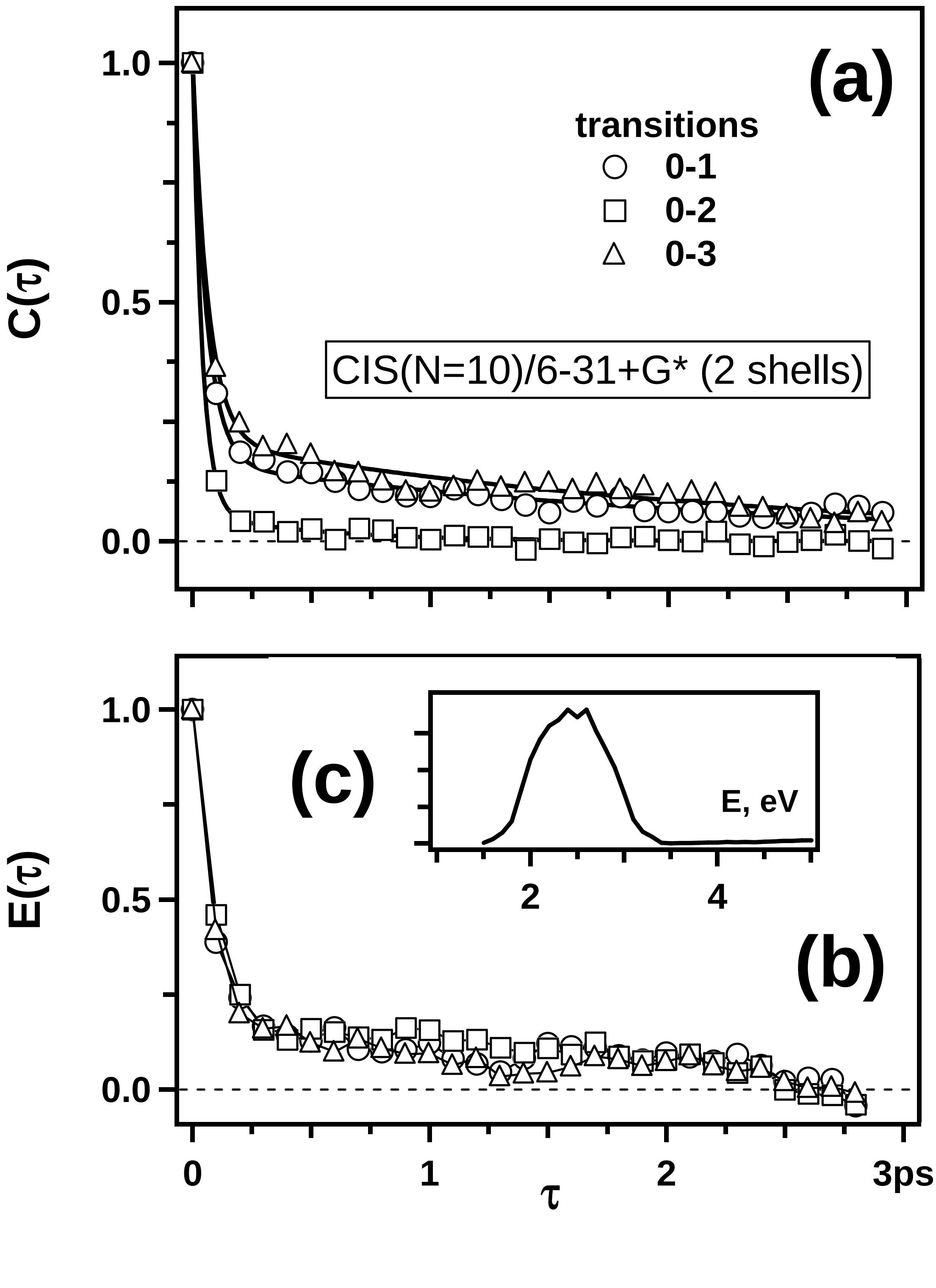

(a)
transitions
0-1
0-2
0-3
CIS(N=10)/6-31+G* (2 shells)
C(τ)
1.0
0.5
0.0
(c)
E, eV
2
4
(b)
E(τ)
1.0
0.5
0.0
0
1
2
3ps
τ

Figure 3S

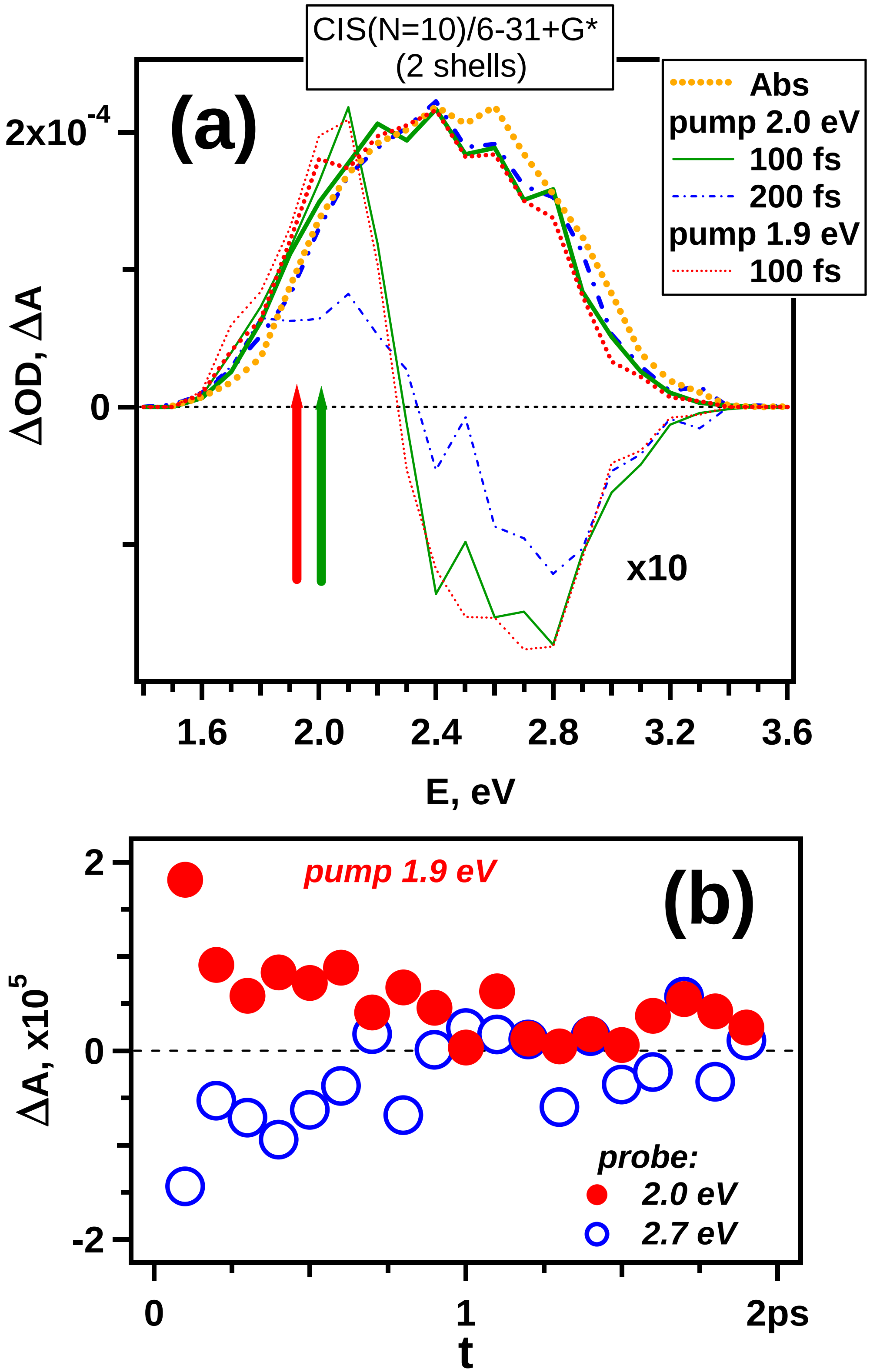
CIS(N=10)/6-31+G*
(2 shells)
(a)
Abs
pump 2.0 eV
100 fs
200 fs
pump 1.9 eV
100 fs
x10
2x10$^{-4}$
0
ΔOD, ΔA
1.6
2.0
2.4
2.8
3.2
3.6
E, eV
(b)
pump 1.9 eV
probe:
2.0 eV
2.7 eV
ΔA, x10$^5$
2
0
-2
0
1
2ps
t